\documentclass[10pt]{article}
\usepackage{graphicx}
\usepackage{epsfig}
\usepackage{color}

\usepackage{amssymb,amsmath}
\usepackage{graphicx}  
\newcommand{\beq}{\begin{equation}}
\newcommand{\eeq}{\end{equation}}
\newcommand{\be}{\begin{equation}}
\newcommand{\ee}{\end{equation}}
\newcommand{\bea}{\begin{eqnarray}}
\newcommand{\eea}{\end{eqnarray}}

\fontencoding{T1}
\fontfamily{garamond}
\fontseries{m}
\fontshape{it}
\fontsize{13}{15}
\selectfont

\begin{document}

\title{Update on Counting Valence Quarks at RHIC}
\author{L.~Maiani$^a$, A.D. Polosa$^b$, V.~Riquer$^b$, C.A.~Salgado$^a$\\
$^a$Dip. Fisica, Universit\`a  di Roma ``La Sapienza'' and INFN, Roma, Italy\\
$^b$INFN, Sezione di Roma, Roma, Italy}
\maketitle
\begin{abstract}
We update our former analysis of the Nuclear Modification Factors (NMF) for different hadron species at RHIC and
LHC. This update is motivated by the new experimental data from STAR which presents differences with the preliminary data used to fix some of the parameters in our model. The main change is the use of AKK fragmentation functions for the hard part of the spectrum and minor adjustments of the coalescence (soft) contribution. We confirm that observation of the NMF for the $f_0$ meson can shed light on its quark composition.
\newline
\newline
{\bf Keywords} Heavy Ion Collisions\newline
{\bf PACS} 12.38.Mh
\end{abstract}

In a recent paper~\cite{Maiani:2006ia} we have studied the $R_{CP}$ and $R_{AA}$ nuclear modification factors for different hadron species at RHIC and LHC using the coalescence model developed in~\cite{Fries:2003vb}. It turns out that these observables are very sensitive to the number of constituent quarks of the hadron species they are calculated for. A striking evidence of this phenomenon is that the $R_{CP}$ is remarkably different for protons and pions, and for $\Lambda$'s and Kaons, being systematically higher for baryons.

Some hadrons, especially the lightest scalar mesons, have controversial interpretations as for their quark structure; see e.g.~\cite{closetorn}. We believe that there are solid reasons to understand the observed sub-GeV scalar meson nonet as four-quark states~\cite{ourpaps}. The same structure may be shared by  the first $0^+$ super-GeV multiplet. Because of the affinity of $f_0$ to Kaons, the more conventional alternative for $f_0$ is $f_0=s\bar s$.

The point made in~\cite{Maiani:2006ia} is that the $R_{CP}$  and $R_{AA}$ observables could indeed be used to discriminate the nature of the $f_0(980)$ ($s\bar s$ meson or $4q$ state). $f_0(980)$ is  a good candidate for such a study since it is one of the mesons that are observable in inclusive reactions at RHIC. 
 
The approach used to describe the production of hadrons at RHIC, in the intermediate momentum range  1.5 GeV$\leq p_T\leq $4 GeV, is a combination of (i) a coalescence model for the soft part of the spectrum and (ii) fragmentation for the hard part. 

Parameters for the coalescence component and the inclusive jet cross sections are fitted to the inclusive production of {\it standard} hadrons, pions, p+$\bar p$, Kaons and $\Lambda$s. Fragmentation functions are taken from deep inelastic processes, typically from Z decays at LEP where large statistics are available.


While the coalescence picture of a multiquark $f_0$ is a straightforward extension of the model, the fragmentation of $f_0$ requires some additional hypothesis on the functional structure of the fragmentation functions.

In~\cite{Maiani:2006ia} we used the fragmentation function set developed by~\cite{vogel} and the following model:
\begin{eqnarray}
&&D_q^{f_0(4q)}(z,Q^2) \sim  0.5(1-z)^{1.5}~\frac{D_q^{\Lambda+\bar \Lambda}(z,Q^2)}{2}\nonumber\\
&&D_g^{f_0(4q)}(z,Q^2) \sim 0.1 (1-z)^{5}~\frac{D_q^{\Lambda+\bar \Lambda}(z,Q^2)}{2}
\end{eqnarray}

In this note, we update our former analysis to take into account the latest data on $R_{AA}$ for $\Lambda$'s and $\Xi$'s~\cite{Abelev:2007xp}.

The new data show qualitative change with respect to the preliminary ones~\cite{Salur:2005nj} used in~\cite{Maiani:2006ia}. In particular, they show that the AKK fragmentation functions describe better the inclusive production of $\Lambda$+$\bar \Lambda$, which is underestimated by the  ones in~\cite{vogel}.

\paragraph{New Fragmentation Functions for $f_0$ and $\Xi$.}In the present analysis we have adopted the AKK fragmentation functions for $\Lambda$+$\bar \Lambda$ and changed the fragmentation model of a 4-quark  meson, in order to keep the agreement with the LEP and NOMAD data on $f_0$ production. Our new parameterization is:

\begin{eqnarray}
&&D_q^{f_0(4q)}(z,Q^2) \sim  0.5~\frac{D_q^{\Lambda+\bar \Lambda}(z,Q^2)}{2}\nonumber\\
&&D_g^{f_0(4q)}(z,Q^2) \sim 0\end{eqnarray}

\begin{figure}[ht]
\begin{center}
\includegraphics[scale=0.8]{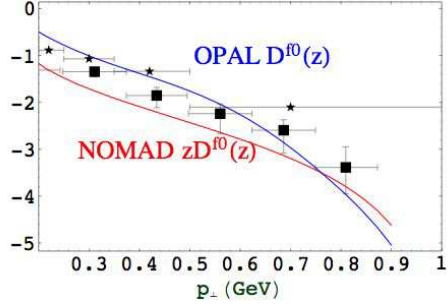}
\end{center}
\vspace{-0.5truecm}
\caption{\footnotesize Comparison with OPAL and NOMAD data on $f_0(980)$. Data from~\cite{opal, nomad}}
\label{fig:ndist}
\end{figure}

The  new fragmentation function is reported in Fig.~\ref{fig:ndist}, blue curve, against the LEP data (stars). The red curve shows $zD_q^{f_0(4q)}(z,Q^2)$ compared to NOMAD data (squares).

For the $\Xi+\bar \Xi$, the new cross sections are reproduced by  fragmentation functions with a suppression factor of $0.5$ with respect to $\Lambda + \bar \Lambda$ but no change in $z\to 1$ behavior: 
\begin{eqnarray}
&&D_q^{\Xi+\bar \Xi}(z,Q^2) \sim  0.5~D_q^{\Lambda+\bar \Lambda}(z,Q^2)\nonumber\\
&&D_g^{\Xi+\bar \Xi}(z,Q^2) \sim 0
\end{eqnarray}

\paragraph{Coalescence parameters.} The new data require a tuning of the coalescence parameters, to fit the reference cross sections. We report in Table~\ref{tableold} below the old parameters, used in~\cite{Maiani:2006ia}.
In our update we have adjusted the quark and antiquark fugacities in Au+Au at RHIC as follows: 
\begin{eqnarray}
&&{\rm Au+Au~@~RHIC} \left\{ \begin{array}{c} \gamma_{u,d,s}=\gamma_{\bar u, \bar d}\\
\hspace{-0.6truecm}\gamma_{\bar s}=0.9 \end{array}\right .
\end{eqnarray}

\begin{figure}[htb!]
\begin{center}
\hspace{-4truecm}
\begin{minipage}[ht]{50mm}
\includegraphics[scale=0.8]{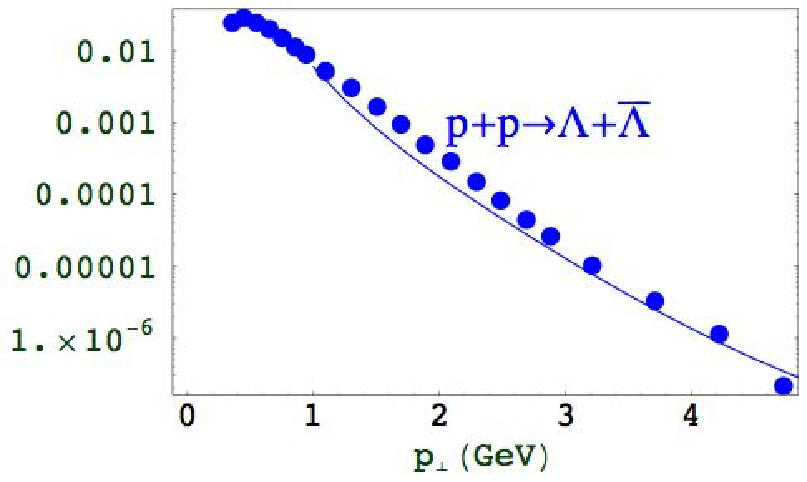}
\end{minipage}
\hspace{4truecm}
\begin{minipage}[ht]{50mm}
\includegraphics[scale=0.8]{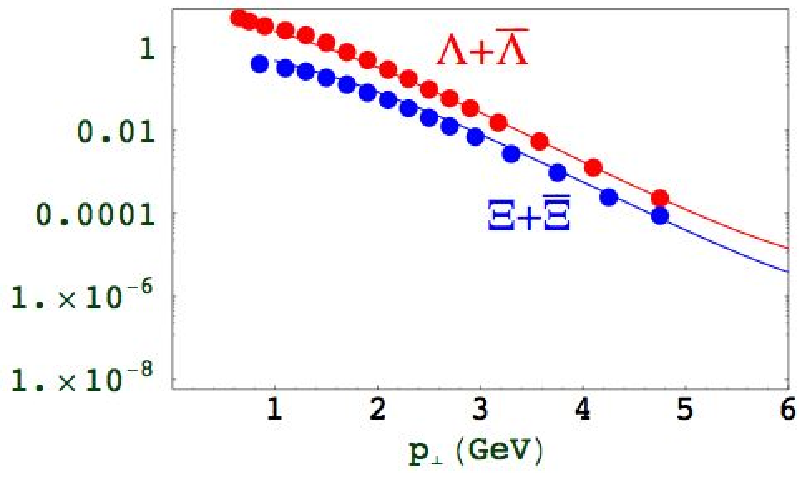}
\end{minipage}
\end{center}
\vspace{-0.5truecm}
\caption{\footnotesize Left panel: Cross sections for production of $\Lambda$'s in proton-proton collisions. Data from~\cite{xsects}. Right panel: Cross sections for central production of $\Lambda$'s and $\Xi$'s in Au+Au collisions at $\sqrt{s_{NN}}=200$~GeV~\cite{Adams:2006ke}. }
\label{calibration}
\end{figure}

The comparison with the new cross sections is summarized in Fig.~\ref{calibration}.

\begin{figure}[ht]
\begin{center}
\hspace{-4truecm}
\begin{minipage}[ht]{50mm}
\includegraphics[scale=0.7]{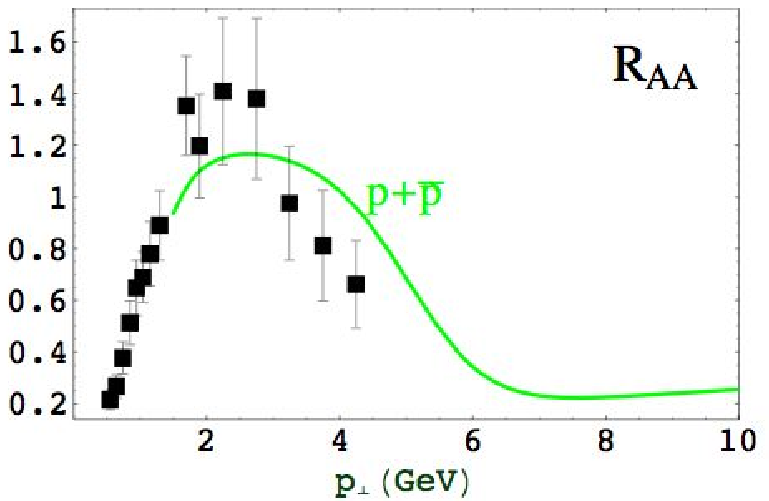}
\end{minipage}
\hspace{3truecm}
\begin{minipage}[ht]{50mm}
\includegraphics[scale=0.7]{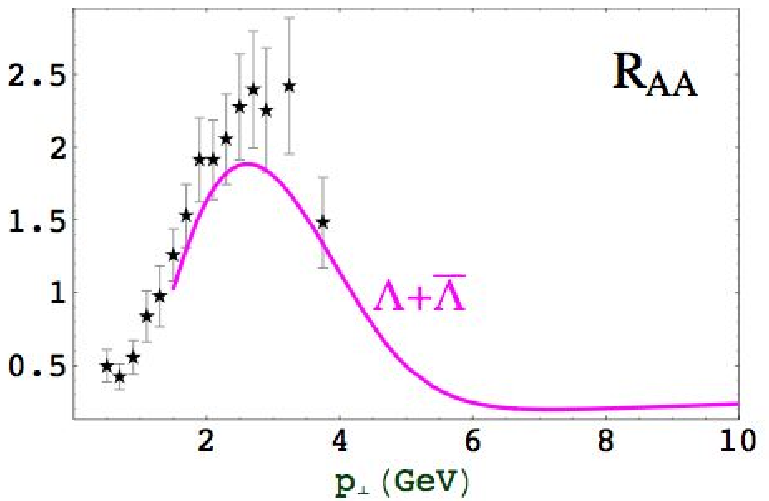}
\end{minipage}
\end{center}
\begin{center}
\hspace{-4truecm}
\begin{minipage}[ht]{50mm}
\includegraphics[scale=0.7]{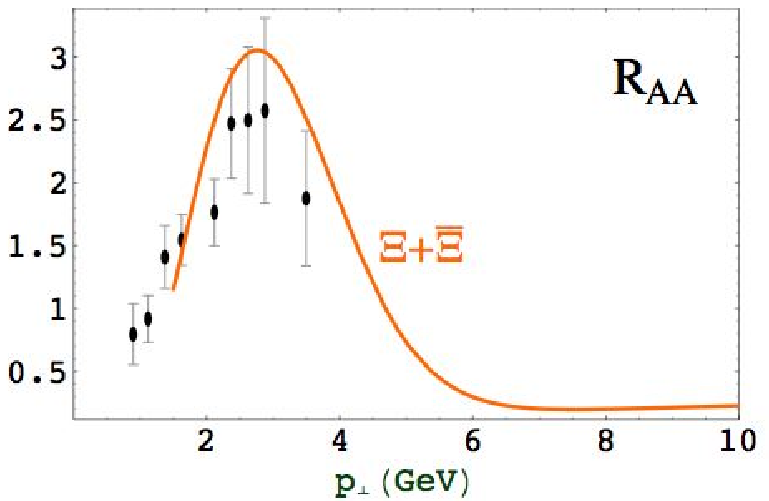}
\end{minipage}
\hspace{3truecm}
\begin{minipage}[ht]{50mm}
\includegraphics[scale=0.7]{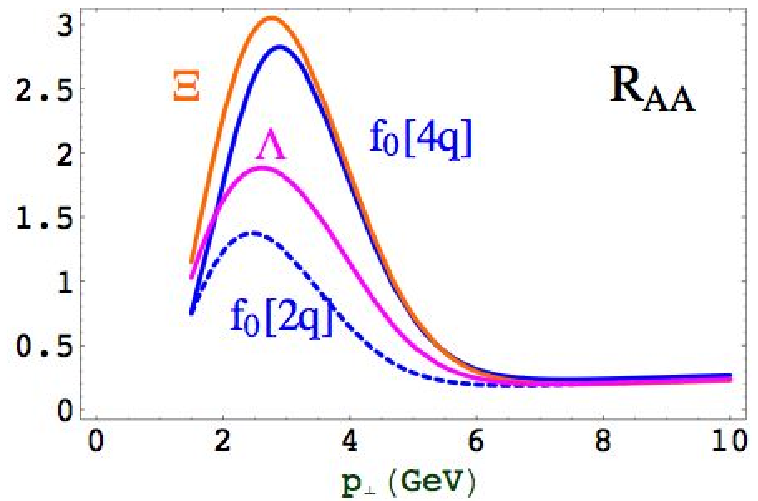}
\end{minipage}
\caption{\footnotesize ${\bf R_{AA}}$ ratios. 
Data from~\cite{Abelev:2007xp}.}
\label{fig:raa}
\end{center}
\end{figure}

\paragraph{Nuclear Modification Factors.} In Fig.~\ref{fig:raa} we show the updated results for $R_{AA}$ for different hadron species.
\newpage
 In the last panel we present our new prediction for the $f_0[4q]$  vs. $f_0[s\bar s]$, compared to the theoretical curves for $\Lambda+\bar \Lambda$ and $\Xi + \bar \Xi$ of Fig.~\ref{fig:raa}.

\begin{figure}[htb!]
\begin{center}
\includegraphics[scale=1]{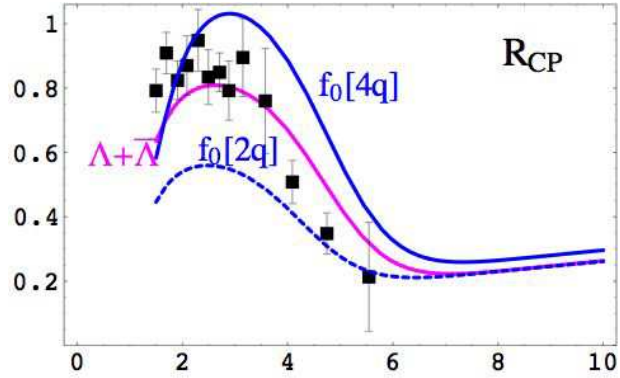}
\end{center}
\vspace{-0.5truecm}
\caption{\footnotesize The ${\bf R_{CP}}$ with the new set of fragmentation functions. Data for $\Lambda$+$\bar \Lambda$ from~\cite{long}.}
\label{fig:newrcp}
\end{figure}

The new fragmentation function set induces a small variation also in the calculation of $R_{CP}$ with respect to the results shown in~\cite{Maiani:2006ia}. We summarize our results for $R_{CP}$ of $f_0$ in Fig.~\ref{fig:newrcp}.

%

\begin{table}[htb]
\caption{{\footnotesize Parameters of the hadron production model in Au+Au @ RHIC, Pb+Pb @LHC and p+p, from Ref.~\cite{Maiani:2006ia}. Numbers in parentheses in the last column refer to peripheral collisions with $b=12$~fm at RHIC and $b=14$~fm at LHC. }}
\begin{center}
\label{tableold}
\begin{tabular}{@{}|c|c|c|c|c|c|c|c|c|}
\hline
~~~~               & $\gamma_{u,d}$ & $\gamma_{\bar u,\bar d}$ & $\gamma_{s,\bar s}$ &$\gamma_{{\rm periph}}$ & $\begin{array}{c} \tau A_T \\ ({\small {\rm fm} ^3)}\end{array}$ &v$_\perp$& $\begin{array}{c}\epsilon_0\\({\rm GeV}^{-1/2})\end{array}$ &N$_{coll}$ \\
\hline
Au+Au @RHIC       	& 1	& 0.9 & 0.8 & 0.7 &  1.27$\cdot 10^3$ & 0.55& 0.82&1146 (26)  \\
\hline
Pb+Pb @LHC          & 1	& 1 & 1 & 0.7 & 11.5$\cdot 10^3$ & 0.68 & 2.5 & 3643(82) \\
\hline
p+p @RHIC(LHC)		&0.4 &  0.4 & 0.12 (0.4)& --  & 13.2 (119) & 0.55 (0.68) & 0& -- \\
\hline
\end{tabular}\\[2pt]
\end{center}
\end{table}


\newpage
\begin{thebibliography}{99}

\bibitem{Maiani:2006ia}
  L.~Maiani, A.~D.~Polosa, V.~Riquer and C.~A.~Salgado,
  Phys.\ Lett.\  B {\bf 645}, 138 (2007)
  [arXiv:hep-ph/0606217].
  
  \bibitem{Fries:2003vb}
  R.~J.~Fries, B.~Muller, C.~Nonaka and S.~A.~Bass,
  Phys.\ Rev.\ Lett.\  {\bf 90}, 202303 (2003)
  [arXiv:nucl-th/0301087].
  
  \bibitem{closetorn}
  F.~E.~Close and N.~A.~Tornqvist,
  J.\ Phys.\ G {\bf 28}, R249 (2002)
  [arXiv:hep-ph/0204205].
  
  \bibitem{ourpaps}
    L.~Maiani, F.~Piccinini, A.~D.~Polosa and V.~Riquer,
  Phys.\ Rev.\ Lett.\  {\bf 93}, 212002 (2004)
  [arXiv:hep-ph/0407017]; 
  Eur.\ Phys.\ J.\  C {\bf 50}, 609 (2007)
  [arXiv:hep-ph/0604018];  L.~Maiani, A.~D.~Polosa and V.~Riquer,
  arXiv:hep-ph/0703272.


\bibitem{vogel}
  D.~de Florian, M.~Stratmann and W.~Vogelsang,
  Phys.\ Rev.\  D {\bf 57}, 5811 (1998)
  [arXiv:hep-ph/9711387].
  
\bibitem{Abelev:2007xp}
  B.~I.~Abelev {\it et al.}  [STAR Collaboration],
  arXiv:0705.2511 [nucl-ex].
  
\bibitem{Salur:2005nj}
  S.~Salur  [STAR Collaboration],
  Nucl.\ Phys.\  A {\bf 774}, 657 (2006)
  [arXiv:nucl-ex/0509036].

 \bibitem{akk}
  S.~Albino, B.~A.~Kniehl and G.~Kramer,
  Nucl.\ Phys.\  B {\bf 725}, 181 (2005)
  [arXiv:hep-ph/0502188];  
  Nucl.\ Phys.\  B {\bf 734}, 50 (2006)
  [arXiv:hep-ph/0510173].
  
\bibitem{opal}
  K.~Ackerstaff {\it et al.}  [OPAL Collaboration],
  Eur.\ Phys.\ J.\  C {\bf 4}, 19 (1998)
  [arXiv:hep-ex/9802013].

\bibitem{nomad}
  P.~Astier {\it et al.}  [NOMAD Collaboration],
  Nucl.\ Phys.\  B {\bf 601} (2001) 3
  [arXiv:hep-ex/0103017].

\bibitem{long}
H.~Long [STAR Collaboration],  J.~Phys. G {\bf 30}, S193 (2004). 


\bibitem{xsects}
  B.~I.~Abelev {\it et al.}  [STAR Collaboration],
  Phys.\ Rev.\  C {\bf 75}, 064901 (2007)
  [arXiv:nucl-ex/0607033].

\bibitem{Adams:2006ke}
  J.~Adams {\it et al.}  [STAR Collaboration],
  Phys.\ Rev.\ Lett.\  {\bf 98}, 062301 (2007)
  [arXiv:nucl-ex/0606014].
  See also {\tt http://www.star.bnl.gov/STAR/all/physicsdatabase/66/data.html}


\end{thebibliography}
\end{document}